\documentclass[final]{aipproc}
\layoutstyle{6x9}
\usepackage{amsmath}
\usepackage{amssymb}
\usepackage{mathtools}
\usepackage{tikz}
\usetikzlibrary{positioning,calc}
\DeclareMathAlphabet{\mathpzc}{OT1}{pzc}{m}{it}
\newcommand{\p}{\textit{p.}}

\DeclareMathOperator{\Amp}{z}
\DeclareMathOperator{\Prob}{\operatorname{Pr}}
\DeclareMathOperator{\Ident}{\mathcal{I}}
\DeclareMathOperator{\sgn}{\operatorname{sgn}}

\renewcommand{\Vec}[1]{\boldsymbol{\mathbf{#1}}}

\newcommand{\abs}[1]{\left|{#1}\right|}

\newcommand{\Z}{\mathbb{Z}}

\newcommand{\Action}{\mathcal{S}}
\newcommand{\Kernel}{\mathcal{K}}

\newcommand{\PathD}[1]{\mathpzc{D}\{#1\}}

\newcommand{\Perms}[1]{\mathbb{S}_{#1}}

\newcommand{\Weight}{\chi}

\newcommand{\Eqref}[1]{Eq.~\eqref{#1}}
\newcommand{\Eqtworef}[2]{Eqs.~\eqref{#1},~\eqref{#2}}
\newcommand{\Figref}[1]{Figure~\ref{#1}}

\newcommand{\Cond}[2]{\left.{#1}\middle|{#2}\right.}
\newcommand{\CondProb}[2]{\Prob\left(\!\Cond{#1}{#2}\!\right)}
\tikzset{forbidden zone/.style n args={4}{shade,shape=circle,inner sep=0pt,minimum size=#1, draw=#2, outer color=#3, inner color=#4}}
\tikzset{forbidden zone/.default={3mm}{black}{black!10}{white}}
\tikzset{point/.style n args={3}{shape=circle,inner sep=0pt,minimum size=#1, draw=#2, fill=#3}}
\tikzset{point/.default={1mm}{black}{white}}
\tikzset{particle/.style={point={3mm}{black}{white}}}
\tikzset{fullparticle/.style={particle,fill=black}}
\tikzset{dotparticle/.style={particle,dotted}}
\tikzset{ordarrow/.style={->}}
\tikzset{dotarrow/.style={->,dotted}}
\begin{document}
\title{Anyons in the Operational Formalism}
\classification{\texttt{03.65.Ca}, \texttt{03.65.Ta}, \texttt{03.65.Vf}}
\author{Klil H. Neori}{
  address={University at Albany, SUNY}
}
\author{Philip Goyal}{
  address={University at Albany, SUNY}
}
\noindent\textit{The following article has been submitted to AIP Conference Proceedings, for the volume dedicated to MaxEnt 2014, which took place Sep 21--26, 2014 in Amboise, France.}

\noindent\textit{After it is published, it will be found at}~\texttt{http://proceedings.aip.org}\textit{.}
\begin{abstract}
\noindent
The operational formalism to quantum mechanics seeks to base the theory on a firm foundation of physically well-motivated axioms~\cite{GoyalKnuthSkilling2010}. It has succeeded in deriving the Feynman rules~\cite{Feynman1948} for general quantum systems. Additional elaborations have applied the same logic to the question of identical particles, confirming the so-called Symmetrization Postulate~\cite{MessiahGreenberg1964}: that the only two options available are fermions and bosons~\cite{NeoriGoyal2013,Goyal2014}. However, this seems to run counter to results in two-dimensional systems, which allow for anyons, particles with statistics which interpolate between Fermi-Dirac and Bose-Einstein~(see~\cite{Lerda1992} for a review).

In this talk we will show that the results in two dimensions can be made compatible with the operational results. That is, we will show that anyonic behavior is a result of the topology of the space in two dimensions~\cite{LaidlawDeWitt1971}, and does not depend on the particles being identical; but that nevertheless, if the particles are identical, the resulting system is still anyonic.

\end{abstract}
\maketitle
\section{What particles are there?}
A dominant thread in physical research has been the search for the basic constituents of Nature. We can trace this thread all the way back to the pre-Socratic philosopher Democritus, who coined the term ``atom''=indivisible, through Gottfried Leibniz and his monads, Rutherford's exploration of the structure of what we now still misleadingly call an atom, and all the way to elementary particle physics, with its Standard Model, one of whose last constituents, the Higgs boson, has been detected at CERN's Large Hadron Collider a few years ago. Current elementary particles only fall into two types: bosons and fermions.

Bosons can be squeezed together endlessly. Photons are bosons, which is why electromagnetic waves are so easily treated classically: the overabundance of photons in the same state masks the quantum nature of light. In modern terms, this property is expressed in their overall wavefunction being symmetric; for example:
\begin{align}
\Psi(x_2,x_1,x_3)&=+\Psi(x_1,x_2,x_3)\text{.}\label{eq:labelexbosons}
\intertext{Fermions, meanwhile, refuse to be in the same state. Electrons are fermions, which is why atoms are built up of progressively filled shells of electrons. In modern terms, this is expressed by their wavefunction being anti-symmetric:}
\Psi(x_2,x_1,x_3)&=-\Psi(x_1,x_2,x_3)\text{.}\label{eq:labelexfermions}
\end{align}

Another important thread, particularly in modern physics, has been the analysis of more complicated systems in terms of particle-like entities, or excitations. For example, under certain conditions, the behavior of solids can be analyzed through phonons, which represent the vibration of the material, and effective electrons and holes, which act somewhat like free electrons or positrons, but take into account the properties of the material from which they stem, which includes many electrons, as well as atomic nuclei. A grand achievement of this point of view was an explanation of low-temperature superconductors in through the use of ``cooper pairs'', pairs of electrons which condense into quasi-particles through an interaction with the atomic lattice. Interfaces between materials or very thin layers may also contain exotic excitations called anyons, which have charges that are a fraction of the elementary, electron charge, and explain some strange behaviors of the quantum Hall effect.
\section{How can one go beyond bosons and fermions, and to what end?}
Phonons and cooper pairs are bosons, while electrons and holes are fermions. None of them challenge the boson-fermion dichotomy. Anyons, however, do, which is somewhat difficult to express. Let us look at the generalization of~\Eqtworef{eq:labelexbosons}{eq:labelexfermions}:
\begin{align}
\Psi(x_2,x_1,x_3)&=\Weight((1\,2))\Psi(x_1,x_2,x_3)\text{;}\label{eq:labelexgeneral}
\intertext{when applied twice we get:}
\Psi(x_1,x_2,x_3)&=\Weight((1\,2))\Psi(x_2,x_1,x_3)=[\Weight((1\,2))]^2\Psi(x_1,x_2,x_3)\text{,}
\end{align}
which seems to indicate that $[\Weight((1\,2))]^2=1$, so that~$\Weight((1\,2))=\pm1$, and only bosons and fermions are allowed. That has ultimately been the basis of several proofs critiqued by~\citet*{MessiahGreenberg1964}, who coined the term~\emph{Symmetrization Postulate} for the assumption that only~\Eqtworef{eq:labelexbosons}{eq:labelexfermions} are possible. In this context it is equivalent to assuming that~$\Weight(\sigma)$, for~$\sigma\in\Perms{N}$, is one-dimensional. They then provided for multi-dimensional alternatives, \emph{paraparticles}.

But~\citet*{Mirman1973} challenged the very basis of considering the change of labels as an operator; after all, other operators refer to actions, such as rotations and translations, that can be performed on a physical system. How do you exchange arbitrary labels? Should there not be a way of expressing particle identity to begin with?

\begin{figure}[ht!]
\centering
\begin{tikzpicture}[scale=1,node distance=5mm,on grid, line cap=round,>=latex]
\begin{scope}
	\coordinate [fullparticle] (a1) at (-0.5cm, 0.5cm);
	\coordinate [particle] (a2) at ($ (a1) + (1cm,0) $);
	\node [left of=a1] {(i)};
	\draw [dashed] (-0.7cm, 0) -- (0.7cm, 0);
	\coordinate [particle] (b1) at (-0.5cm, -0.5cm);
	\coordinate [fullparticle] (b2) at ($ (b1)+ (1cm,0) $);
	\node [left of=b1] {(ii)};
	\coordinate [fullparticle] (c1) at (2cm, 0);
	\coordinate [fullparticle] (c2) at ($ (c1) + (1cm,0)$);
	\coordinate [fullparticle,opacity=0] (cmid) at ($ (c1)!0.8!(c2) $);
	\node [below of=cmid] {(iii)};
	\draw [dotarrow] ($ (a2) + (0.5cm, 0) $) to ($ (c1) + (-0.5cm, 0.25cm) $);
	\draw [dotarrow] ($ (b2) + (0.5cm, 0) $) to ($ (c1) + (-0.5cm,-0.25cm) $);
	\draw [dashed] (3.5cm, 0.7cm) -- (3.5cm, -0.7cm);
	\coordinate [fullparticle] (d1) at (4.5cm, 0);
	\coordinate [fullparticle] (d2) at ($ (d1) + (1.5cm, 0) $);
	\coordinate [dotparticle] (dmid) at ($ (d1)!0.5!(d2) $);
	\draw [dotarrow] (d1) -- (dmid);
	\draw [dotarrow] (d2) -- (dmid);
	\draw [thick] ($ (dmid) + (-0.3cm, 0.2cm) $) -- ($ (dmid) + (0.3cm, -0.2cm) $);
	\draw [thick] ($ (dmid) + (-0.3cm, -0.2cm) $) -- ($ (dmid) + (0.3cm, 0.2cm) $);
	\node [left of=d1] {(iv)};
\end{scope}
\end{tikzpicture}
\caption{Two distinguishable particles in two distinct states, (i) and (ii), become two identical particles in a single state (iii) after reduction. Coincidence (iv) is forbidden.}\label{fig:symmandnocoin}
\end{figure}
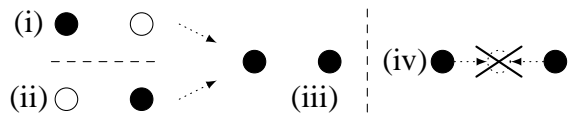

In response to this, and with an eye towards the Gibbs paradox,~\citet*{LeinaasMyrheim1977} independently%
\footnote{The authors were apparently unaware of~\citet*{Souriau1967} and~\citet*{LaidlawDeWitt1971}, which also introduced a reduced configuration space, albeit using different quantization schemes.}%
\ introduced the notion of reducing the configuration space of the classical system which is to be quantized, so that it already reflects the identity of the particles~(See~\Figref{fig:symmandnocoin}(i)--(iii)).
 
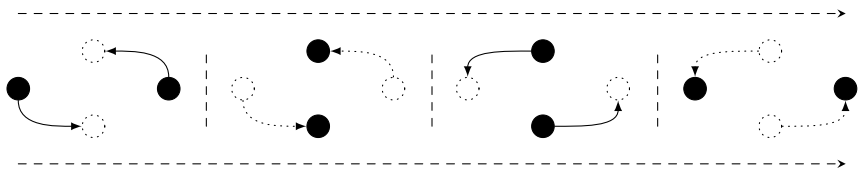
\begin{figure}[ht!]
\centering
\begin{tikzpicture}[scale=1,node distance=5mm,on grid, line cap=round,>=latex]
\foreach \xshiftouter/\firstlineoutangle/\xstart/\ystart/\xfinish/\yfinish in {0cm/90/1/0/0/0.5, 6cm/180/0/0.5/-1/0} {
	\foreach \xshiftinner/\startstyle/\midstyle/\linestyle in {0cm/fullparticle/dotparticle/ordarrow, 3cm/dotparticle/fullparticle/dotarrow} {
	\pgfmathtruncatemacro{\xshifttotal}{\xshiftouter+\xshiftinner};
	\begin{scope}[xshift=\xshifttotal]
		\coordinate [\startstyle] (st) at (\xstart cm, \ystart cm);
		\coordinate [\startstyle] (-st) at ($ {-1}*(st) $);
		\coordinate [\midstyle] (fi) at (\xfinish cm, \yfinish cm);
		\coordinate [\midstyle] (-fi) at ($ {-1}*(fi) $);
		\begin{scope}[\linestyle]
			\draw [out=\firstlineoutangle, in={\firstlineoutangle-90}] (st) to (fi);
			\draw [out={\firstlineoutangle+180}, in={\firstlineoutangle+90}] (-st) to (-fi);
		\end{scope}
	\end{scope}
	}
}
\foreach \xline in {0, 3, 6}
	\draw [dashed] ({\xline cm+1.5cm}, -0.5cm) -- ({\xline cm+1.5cm}, +0.5cm);
\foreach \yarrow in {-1, 1}
	\draw[dashed,->,>=stealth] (-1cm,\yarrow cm) to (10cm, \yarrow cm);
\end{tikzpicture}
\caption{Exchange accrues a phase of~$e^{i\varphi}$; $\varphi=0$ for bosons, $\pi$ for fermions, general for anyons.}\label{fig:illpartexch}
\end{figure}

Additionally, coincidence is forbidden~(See~\Figref{fig:symmandnocoin}(iv)). This is essential so that the resulting space remains a differentiable manifold, which is necessary for quantization, although this is only true for three or more dimensions, as noted in~\citet*{Bloore1979}. In fact, it is this restriction that creates the interesting topological phenomena in two dimensions, even though the removal is unnecessary there; without it, only bosons are possible under this framework. Nevertheless, once the incidence points have been removed and the space reduced, the authors pursue the notion of an operator corresponding to ``moving'' the wavefunction around and returning it to the same point. In essence, they present generalizing the likes of~\Eqref{eq:labelexgeneral} to the form:
\begin{equation}
\Psi(x_2,x_1,x_3)=\Weight(\textsl{path exchanging }x_1\textsl{ and }x_2)\Psi(x_1,x_2,x_3)\text{;}
\end{equation}
it is in the representation of these different paths that the topological degree of freedom manifests; in two dimensions, this leads to anyons, a term coined by~\citet*{Wilczek1982b}. Furthermore, bosons and fermions are still acceptable as anyons. Therefore, if we wish to illustrate the situation more visually, we could discuss a phase that is accrued by moving particles around each other once counterclockwise, it being either~$1$ for bosons,~$-1$ for fermions, or generally~$e^{i\varphi}$ for anyons, which are seen to interpolate between the two extremes~(see~\Figref{fig:illpartexch}).

The discussion so far should make it clear that wavefunctions are not a comfortable setting in which to discuss the topological degree of freedom represented by anyonic behavior. Indeed, we will find that path integrals are a far better tool for the job, as was found by~\citet*{Schulman1967,Schulman1968,Schulman1971} and in a more general way, including the identical-particle case, by~\citet*{LaidlawDeWitt1971}.

More importantly, in previous work we have produced a result concerning identical particles using the Feynman rules, of which the path integrals are the continuous limit. However, that result challenges our assertion to be able to handle anyons: it seems to rule them out entirely. We turn to it next, before expanding upon paths in topology, and our resolution of this conundrum.
\section{What does the operational formalism say about identical particles?}
\begin{figure}[ht!]
\centering
\begin{tikzpicture}[scale=1,node distance=5mm,on grid, line cap=round,>=latex]
\def\anglestart{-180}
\def\rstart{0.5}
\def\anglefinish{45}
\def\rfinish{0.7}
\foreach \xshifttotal/\startstyle/\finstyle/\linestyle in {0cm/fullparticle/dotparticle/ordarrow, 3cm/dotparticle/fullparticle/dotarrow} {
\begin{scope}[xshift=\xshifttotal]
	\coordinate [\startstyle,label=left:$\Vec{a}$] (st) at (\anglestart:\rstart cm);
	\coordinate [\finstyle,label=right:$\Vec{b}$] (fi) at (\anglefinish:\rfinish cm);
	\begin{scope}[\linestyle]
		\draw (st) to (fi);
	\end{scope}
\end{scope}
}
\foreach \xline in {0}
{
	\begin{scope}[xshift=\xline]
		\draw [dashed] (1.5cm, -0.5cm) -- (1.5cm, +0.5cm);
	\end{scope}	
}
\foreach \yarrow in {-1, 1}
	\draw[dashed,->,>=stealth] (-1cm,\yarrow cm) to (4cm, \yarrow cm);
\node at (1.2cm, -1.5cm) {(i) $\Amp(\Vec{a}\to\Vec{b})$};
\begin{scope}[xshift=6.5cm]
	\coordinate [fullparticle,label=left:$\Vec{a}$] (st) at (\anglestart:\rstart cm);
	\foreach \midfactor/\midname/\labelplace/\labelcontents in {0.3/midc/left/\text{$\Vec{c}$}, 0.7/midd/right/\text{$\Vec{d}$}} {
		\pgfmathsetmacro{\rcurrent}{(1-\midfactor)*\rstart+\midfactor*\rfinish};
		\pgfmathsetmacro{\anglecurrent}{(1-\midfactor)*\anglestart+\midfactor*\anglefinish};
		\coordinate [dotparticle, label=\labelplace:\labelcontents] (\midname) at (\anglecurrent:\rcurrent cm);
	}
	\coordinate [dotparticle,label=right:$\Vec{b}$] (fi) at (\anglefinish:\rfinish cm);
	\foreach \currpoint/\nextpoint/\linestyle in {st/midc/ordarrow, midc/midd/dotarrow, midd/fi/dotarrow}
		\draw [\linestyle] (\currpoint) to (\nextpoint);
	\node at (0cm, -1.5cm) {(ii) $\Amp(\Vec{a}\to\Vec{c}\to\Vec{d}\to\Vec{b})$};
\end{scope}
\begin{scope}[xshift=11cm]
	\def\endindex{7}
	\pgfmathtruncatemacro{\penultindex}{\endindex-1};
	\coordinate [fullparticle,label=left:$\Vec{a}$] (pt0) at (\anglestart:\rstart cm);
	\coordinate [dotparticle,label=right:$\Vec{b}$] (pt\endindex) at (\anglefinish:\rfinish cm);
	\foreach \forwardindex in {1,...,\penultindex} 
	{
		\pgfmathtruncatemacro{\midindex}{\endindex-\forwardindex};
		\pgfmathsetmacro{\midfactor}{\midindex/\endindex};
		\pgfmathsetmacro{\rcurrent}{(1-\midfactor)*\rstart+\midfactor*\rfinish};
		\pgfmathsetmacro{\anglecurrent}{(1-\midfactor)*\anglestart+\midfactor*\anglefinish};
		\coordinate [dotparticle] (pt\midindex) at (\anglecurrent:\rcurrent cm);
		\pgfmathtruncatemacro{\nextindex}{\midindex+1};
		\draw [dotarrow] (pt\midindex.center) to (pt\nextindex.center);
	}
	\draw [ordarrow] (pt0) to (pt1.center);
	\node at (0cm, -1.5cm) {(iii) $\Amp(\Vec{a}\to\dotsm\to\Vec{b})$};
\end{scope}
\end{tikzpicture}
\caption{From transitions to paths with corresponding amplitudes.}\label{fig:transitiontopath}
\end{figure}
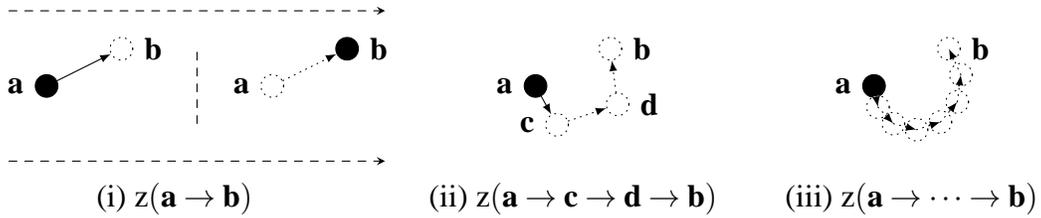

The operational formalism starts out with a physical system, subject to a succession of measurements with potential outcomes~$\{a, b,\dotsc\}$. These are strung together into series of outcomes,~$a\to b\to\dotsm$. The results of~\citet*{GoyalKnuthSkilling2010} show that a set of well-motivated axioms is enough to require that probability amplitudes~$\Amp(a\to b\to \dotsm)$ satisfy the Feynman rules~(as presented in~\citet*{Feynman1948}):
\begin{align}
&\Amp(a\to b\to c)=\Amp(a\to b)\Amp(b\to c)\\
&\Amp(a\to \{b\textsl{ or }c\}\to d)=\Amp(a\to b\to d)+\Amp(a\to c\to d)\\
&\CondProb{b\to\dotsm}{a}=\abs{\Amp(a\to b\to \dotsm)}^2\text{,}
\end{align}
where~$\CondProb{b\to\dotsm}{a}$ is the conditional probability of the rest of the outcomes given that~$a$ was the first. Furthermore, if the outcome space is physical space, and if we analyze transitions with more and more measurements, which are closer to each other both in space and time, the amplitudes for series of outcomes become, in the~$\Delta t\to0$ limit, amplitudes or phases for individual paths, which feature in path integrals~(see~\Figref{fig:transitiontopath}).

\begin{figure}[ht!]
\centering
\begin{tikzpicture}[scale=1,node distance=5mm,on grid, line cap=round,>=latex]
\foreach \sublabel/\xshiftouter/\xstart/\ystart/\xend/\yend in {\text{(i) Direct:~$\alpha_{\textsl{dir}}$}/0cm/0.5cm/-0.5cm/0.5cm/0.5cm, \text{(ii) Opposite:~$\alpha_{\textsl{op}}$}/5cm/0.5cm/-0.5cm/-0.5cm/0.5cm} 
{
	\begin{scope}[xshift=\xshiftouter]
		\foreach \xshiftinner/\startstyle/\linestyle/\endstyle in {0/fullparticle/ordarrow/dotparticle, 2cm/dotparticle/dotarrow/fullparticle} 
		{
			\begin{scope}[xshift=\xshiftinner]
				\coordinate [\startstyle] (sta) at (\xstart, \ystart);
				\coordinate [\startstyle] (stb) at (-\xstart, \ystart);
				\coordinate [\endstyle] (enda) at (\xend, \yend);
				\coordinate [\endstyle] (endb) at (-\xend, \yend);
				\begin{scope}[\linestyle]
					\draw (sta) to (enda);
					\draw (stb) to (endb);
				\end{scope}
			\end{scope}
		}
		\draw [dashed] (1cm, -0.5cm) to (1cm, 0.5cm);
		\foreach \yarrow in {-1, 1}
			\draw[dashed,->,>=stealth] (-1cm,\yarrow cm) to (3cm, \yarrow cm);
		\node at (1cm, -1.5cm) {\sublabel};
	\end{scope}
}
\end{tikzpicture}
\caption{%
Two transitions with same measured result. $\Amp_{\textsl{total}}=\alpha_{\textsl{dir}}\pm\alpha_{\textsl{op}}$}
\label{fig:twotransitions}
\end{figure}
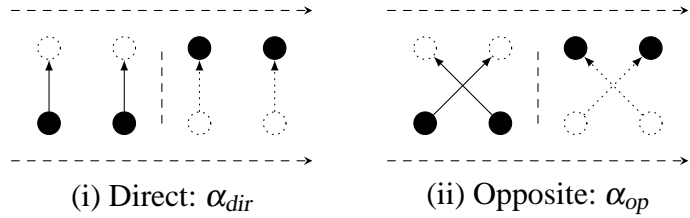

The operational approach was applied to the question of identical particles by~\citet*{NeoriGoyal2013} and~\citet*{Goyal2014}. The results only allowed for two types of solutions: if~$\alpha_{\sigma}$ is the transition amplitude for a system of~$N$ distinguishable particles from a given initial state to a permutation~$\sigma\in\Perms{N}$ of a given final state~(see~\Figref{fig:twotransitions}), then either:
\begin{equation}
z_{\textsl{total}}=\sum\limits_{\sigma}\alpha_{\sigma}\quad\text{or}\quad z_{\textsl{total}}=\sum\limits_{\sigma}\sgn(\sigma)\alpha_{\sigma}\label{eq:opbosfers}\text{,}
\end{equation}
yielding~\emph{operational bosons} or~\emph{operational fermions}, respectively, where~$\sgn(\sigma)$ is~$1$ for permutations made up of an even number of transpositions,~$-1$ otherwise. For two particles, we specialize to:
\begin{equation}
z_{\textsl{total}}=\alpha_{\Ident}\pm\alpha_{(1\,2)}=\alpha_{\textsl{dir}}\pm\alpha_{\textsl{op}}\text{;}\label{eq:optwoparts}
\end{equation}
note that for fermions, the choice of which distinguishable particle transition is ``direct'' and which is ``opposite'' is arbitrary. We will return to this later.

On the surface, it seems that this formalism leaves no room for anyons. However, this result applies regardless of what the amplitudes~$\alpha_{\sigma}$, of the distinguishable-particle systems, are. They are treated as black boxes, so behavior in individual paths of exchange has no influence here. It is in these black boxes that anyonic behavior is hiding. In order to see why that is, it is essential to realize a commonly ignored aspect of anyonic behavior: that it does not depend on the identity of the particles, but only on the fact that they reside in two dimensions, and that they cannot coincide in space%
\footnote{The earliest references we have found to this insight are in~\citet*{Dowker1985} and~\citet*{GoldinMenikoffSharp1985}.}%
. One way to understand this restriction, as argued in~\citet*{Wilczek1990} (\p~16), is that a limitless reduction in distance would require the introduction of unlimited possibilities for particles created by the high energies involved, which would take us far away from the system we wish to analyze. Let us now turn to the properties of these anyonic black boxes, by rejoining our discussion of topology and exchange paths.
\section{What does topology say about exchange?}
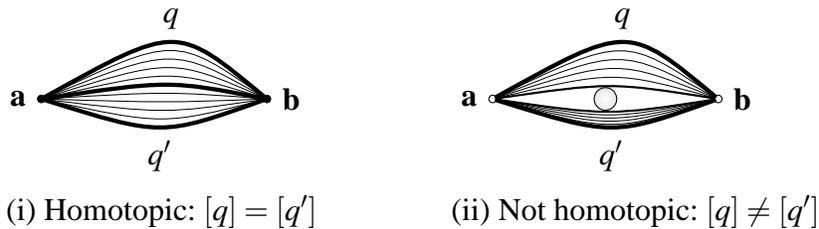
\begin{figure}[ht!]
\centering
\begin{tikzpicture}[scale=1,node distance=5mm,on grid, line cap=round,>=latex]
\begin{scope}
	\node at (0.1cm, -1.5cm) {(i) Homotopic: $[q]=[q']$};
	\coordinate [point, fill=black, label=left:$\Vec{a}$] (a) at (-1.5cm, 0);
	\coordinate [point, fill=black, label=right:$\Vec{b}$] (b) at (1.5cm, 0);
	\coordinate (top) at (3mm, 10mm);
	\coordinate (bottom) at (1mm, -5mm);
	\foreach \l in {0, 0.1,...,1}
		\draw[ultra thin] (a) .. controls ($ \l*(bottom) + {1-\l}*(top) $) .. (b);
	\draw[ultra thick] (a) .. controls (top) .. node[above] {$q$} (b);
	\draw[ultra thick] (a) .. controls (bottom) .. node[below] {$q'$} (b);
	\draw[ultra thick] (a) .. controls ($ 0.5*(bottom) + 0.5*(top) $) .. (b);
\end{scope}
\begin{scope}[xshift=6cm]
	\node at (0.4cm, -1.5cm) {(ii) Not homotopic: $[q]\neq[q']$};
	\coordinate [point, label=left:$\Vec{a}$] (a) at (-1.5cm, 0);
	\coordinate [point, label=right:$\Vec{b}$] (b) at (1.5cm, 0);
	\coordinate (top) at (3mm, 10mm);
	\coordinate (bottom) at (1mm, -5mm);
	\coordinate [forbidden zone] (hole) at (0,0);
	\coordinate (topforbid) at ($ (hole)!0.225!(top) $);
	\coordinate (botforbid) at ($ (hole)!0.45!(bottom) $);
	\foreach \pathcon/\edgecon in {top/topforbid, bottom/botforbid}
		\foreach \l in {0, 0.2,...,1}
			\draw[ultra thin] (a) .. controls ($ (\pathcon)!\l!(\edgecon) $) .. (b);
	\draw[ultra thick] (a) .. controls (top) .. node[above] {$q$} (b);
	\draw[ultra thick] (a) .. controls (bottom) .. node[below] {$q'$} (b);
	\draw[thick] (a) .. controls (topforbid) .. (b);
	\draw[thick] (a) .. controls (botforbid) .. (b);
\end{scope}
\end{tikzpicture}
\caption{The path~$q$ (i) can or (ii) cannot be continuously deformed into~$q'$.}\label{fig:simpvsmultconnected}
\end{figure}

In, \emph{multiply-connected spaces}, not all loops can be smoothly contracted into a point, or, equivalently, not all paths between a given pair of points can be smoothly deformed into each other. Paths between the same two points (or loops starting and ending at the same point) which can be smoothly deformed into one another are called~\emph{homotopic}. This is a an equivalence relation which splits the set of paths between two points into~\emph{homotopy classes}; for a path~$q$, its homotopy class is~$[q]$, and two paths are homotopic if and only if they are in the same class, or~$[q]=[q']$ (as in~\Figref{fig:simpvsmultconnected}(i)); otherwise, the paths are not homotopic, which we can write as~$[q]\neq[q']$ (as in~\Figref{fig:simpvsmultconnected}(ii)).

The Aharonov-Bohm effect~\cite{AharonovBohm1959} was the first to take into account that quantum mechanics behaves strangely in multiply-connected spaces, although in their case, the resulting degree of freedom came about due to a physical source outside the space of electrons, namely the flux of the confined solenoid, instead of being intrinsic. Much like multi-valued functions, this effect hinges on the space being multiply-connected.

Although it was not originally presented in this way, it is easiest to understand this phenomenon as a topological phase accruing differently to paths in the integral,when those are in different homotopy classes, and thus creating novel interference effects, depending on the flux in the confined solenoid:
\begin{align}
\Kernel({\Vec{b}}, t_{\Vec{b}}\,; {\Vec{a}}, t_{\Vec{a}})&=\int\PathD{\Vec{x}(t)}\exp\left(i\Action\{\Vec{x}(t)\}/\hbar\right)=\notag\\
&=\smashoperator[r]{\sum\limits_{[q]}}\Weight([\Vec{x}(t)])\Kernel^{[q]}({\Vec{b}}, t_{\Vec{b}}\,; {\Vec{a}}, t_{\Vec{a}})\label{eq:pathint}\text{,}
\intertext{where~$\Action\{\Vec{x}(t)\}$ is the action functional,~$[q]$ indexes the homotopy classes, and:}
\Kernel^{[q]}({\Vec{b}}, t_{\Vec{b}}\,; {\Vec{a}}, t_{\Vec{a}})&\triangleq\int\PathD{\Vec{x}(t): [\Vec{x}(t)]=[q]}\exp\left(i\Action\{\Vec{x}(t)\}/\hbar\right)\text{,}
\end{align}
that is, this is an integral solely over the paths belonging to~$[q]$. This is equivalent to the effect of the singular vector potential, as depicted in~\citet*{AharonovBohm1959}.
\subsection{How do we get distinguishable anyons?}
\citet*{Schulman1967,Schulman1968} discovered how to treat multiply-connected spaces using Feynman path integrals, leading to~\Eqref{eq:pathint}, and~\citet*{LaidlawDeWitt1971} applied it to the case of identical particles, albeit only in three dimensions.

We have improved upon their argument in order to analyze more general spaces. Instead of rooting the analysis in the fundamental~\emph{group}, that is, a group of homotopy classes of paths starting and ending at the same arbitrary point, we made use of the fundamental~\emph{groupoid}, which is made up of all homotopy classes in the space. This allows us to choose the topological phase of paths in a way that is compatible with the topological degrees of freedom, but also directly implements the space's symmetry.  This leads directly to distinguishable anyons. 

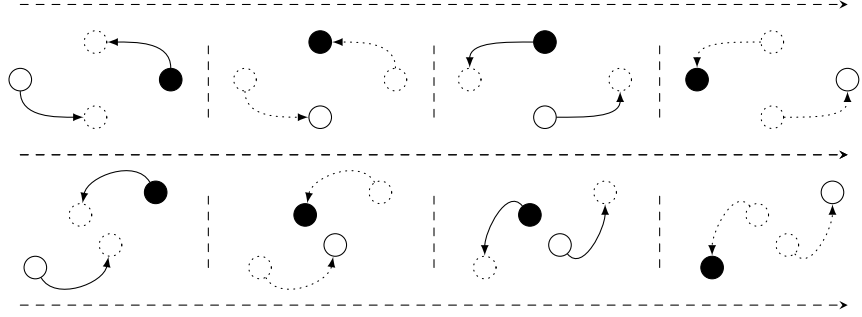
\begin{figure}[ht!]
\centering
\begin{tikzpicture}[scale=1,node distance=5mm,on grid, line cap=round,>=latex]
\begin{scope}
\foreach \xshiftouter/\firstlineoutangle/\xstart/\ystart/\xfinish/\yfinish in {0cm/90/1/0/0/0.5, 6cm/180/0/0.5/-1/0} {
	\foreach \xshiftinner/\startstylea/\startstyleb/\midstylea/\midstyleb/\linestyle in {0cm/fullparticle/particle/dotparticle/dotparticle/ordarrow, 3cm/dotparticle/dotparticle/fullparticle/particle/dotarrow} {
	\pgfmathtruncatemacro{\xshifttotal}{\xshiftouter+\xshiftinner};
	\begin{scope}[xshift=\xshifttotal]
		\coordinate [\startstylea] (st) at (\xstart cm, \ystart cm);
		\coordinate [\startstyleb] (-st) at ($ {-1}*(st) $);
		\coordinate [\midstylea] (fi) at (\xfinish cm, \yfinish cm);
		\coordinate [\midstyleb] (-fi) at ($ {-1}*(fi) $);
		\begin{scope}[\linestyle]
			\draw [out=\firstlineoutangle, in={\firstlineoutangle-90}] (st) to (fi);
			\draw [out={\firstlineoutangle+180}, in={\firstlineoutangle+90}] (-st) to (-fi);
		\end{scope}
	\end{scope}
	}
}
\foreach \xline in {0, 3, 6}
	\draw [dashed] ({\xline cm+1.5cm}, -0.5cm) -- ({\xline cm+1.5cm}, +0.5cm);
\foreach \yarrow in {-1, 1}
	\draw[dashed,->,>=stealth] (-1cm,\yarrow cm) to (10cm, \yarrow cm);
\end{scope}
\begin{scope}[yshift=-2cm]
\foreach \xshiftouter/\firstlineoutangle/\firstlineinangle/\xstart/\ystart/\xfinish/\yfinish in {0cm/120/80/0.8/0.5/-0.2/0.2, 6cm/130/90/-0.2/0.2/-0.8/-0.5} {
	\foreach \xshiftinner/\startstylea/\startstyleb/\midstylea/\midstyleb/\linestyle in {0cm/fullparticle/particle/dotparticle/dotparticle/ordarrow, 3cm/dotparticle/dotparticle/fullparticle/particle/dotarrow} {
	\pgfmathtruncatemacro{\xshifttotal}{\xshiftouter+\xshiftinner};
	\begin{scope}[xshift=\xshifttotal]
		\coordinate [\startstylea] (st) at (\xstart cm, \ystart cm);
		\coordinate [\startstyleb] (-st) at ($ {-1}*(st) $);
		\coordinate [\midstylea] (fi) at (\xfinish cm, \yfinish cm);
		\coordinate [\midstyleb] (-fi) at ($ {-1}*(fi) $);
		\begin{scope}[\linestyle]
			\draw [out=\firstlineoutangle, in=\firstlineinangle] (st) to (fi);
			\draw [out={\firstlineoutangle+180}, in={\firstlineinangle+180}] (-st) to (-fi);
		\end{scope}
	\end{scope}
	}
}
\foreach \xline in {0, 3, 6}
	\draw [dashed] ({\xline cm+1.5cm}, -0.5cm) -- ({\xline cm+1.5cm}, +0.5cm);
\foreach \yarrow in {-1, 1}
	\draw[dashed,->,>=stealth] (-1cm,\yarrow cm) to (10cm, \yarrow cm);
\end{scope}
\end{tikzpicture}
\caption{Both these exchanges of distinguishable anyons accrue $e^{i\varphi}$, as in \Figref{fig:illpartexch}.}\label{fig:allexchanges}
\end{figure}

For example, the fundamental group of the space of two particles in two dimensions which cannot coincide is isomorphic to~$\Z$ under addition, with~$1$, its generator, corresponding to a complete counter-clockwise rotation, with corresponding phase~$e^{i\theta}$. The topological phase of paths can then be chosen so that any simple, counter-clockwise exchange, or half-rotation, would accrue~$e^{i\varphi}=e^{i\theta/2}$. This leads to anyonic behavior without requiring that the particles be identical~(see~\Figref{fig:allexchanges}).
\section{How can the two be combined?}
\begin{figure}[ht!]
\centering
\begin{tikzpicture}[scale=1,node distance=5mm,on grid, line cap=round,>=latex]
\def\anglestart{-180}
\def\rstart{1}
\begin{scope}[xshift=0]
	\def\anglefinish{0}
	\def\rfinish{1}
	\def\endindex{10}
	\pgfmathtruncatemacro{\penultindex}{\endindex-1};
	\coordinate [fullparticle,label=left:$1$] (pt0) at (\anglestart:\rstart cm);
	\coordinate [fullparticle,label=right:$2$] (ref0) at ($ {-1}*(pt0) $);
	\coordinate [dotparticle,opacity=0] (pt\endindex) at (\anglefinish:\rfinish cm);
	\coordinate [dotparticle,opacity=0] (ref\endindex) at ($ {-1}*(pt\endindex) $);
	\foreach \forwardindex in {1,...,\penultindex} 
	{
		\pgfmathtruncatemacro{\midindex}{\endindex-\forwardindex};
		\pgfmathsetmacro{\midfactor}{\midindex/\endindex};
		\pgfmathsetmacro{\rcurrent}{(1-\midfactor)*\rstart+\midfactor*\rfinish};
		\pgfmathsetmacro{\anglecurrent}{(1-\midfactor)*\anglestart+\midfactor*\anglefinish};
		\coordinate [dotparticle] (pt\midindex) at (\anglecurrent:\rcurrent cm);
		\coordinate [dotparticle] (ref\midindex) at ($ {-1}*(pt\midindex) $);
		\foreach \pointtype in {pt, ref} 
		{
			\pgfmathtruncatemacro{\nextindex}{\midindex+1};
			\draw [dotarrow] (\pointtype\midindex.center) to (\pointtype\nextindex.center);
		};
	};
	\foreach \pointtype in {pt, ref} 
		\draw [ordarrow] (\pointtype0) to (\pointtype1.center);
	\node at (0cm, -1.5cm) {(i) Exchange path};
\end{scope}
\begin{scope}[xshift=4.5cm]
	\def\anglefinish{-162}
	\def\rfinish{1}
	\def\endindex{1}
	\coordinate [fullparticle,label=left:$1$] (pt0) at (\anglestart:\rstart cm);
	\coordinate [fullparticle,label=right:$2$] (ref0) at ($ {-1}*(pt0) $);
	\coordinate [dotparticle] (pt\endindex) at (\anglefinish:\rfinish cm);
	\coordinate [dotparticle] (ref\endindex) at ($ {-1}*(pt\endindex) $);
	\foreach \pointtype in {pt, ref} 
		\draw [ordarrow] (\pointtype0) to (\pointtype1.center);
	\node at (0cm, -1.5cm) {(ii) Small step direct};
\end{scope}
\begin{scope}[xshift=9cm]
	\def\anglefinish{18}
	\def\rfinish{1}
	\def\endindex{1}
	\coordinate [fullparticle,label=left:$1$] (pt0) at (\anglestart:\rstart cm);
	\coordinate [fullparticle,label=right:$2$] (ref0) at ($ {-1}*(pt0) $);
	\coordinate [dotparticle] (pt\endindex) at (\anglefinish:\rfinish cm);
	\coordinate [dotparticle] (ref\endindex) at ($ {-1}*(pt\endindex) $);
	\foreach \pointtype in {pt, ref} 
		\draw [ordarrow] (\pointtype0) to (\pointtype1.center);
	\node at (0cm, -1.5cm) {(iii) Small step opposite};
\end{scope}
\end{tikzpicture}
\caption{Exchange path with illustration of direct and opposite transitions.}\label{fig:exchangepath}
\end{figure}
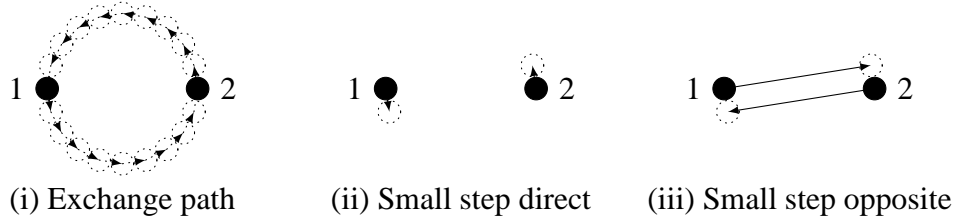

Returning to the operational result in~\Eqref{eq:optwoparts}, and the move from transitions to paths in~\Figref{fig:transitiontopath}, suppose we take an exchange of identical particles as in~\Figref{fig:illpartexch}, and approach the phase of a path by increasing the number of intermediary points between the initial and final configuration, separated by decreasing time intervals~$\Delta t$:
\begin{equation}
\Amp_{\textsl{exch}}=\Amp(\Vec{x}_0\to \Vec{x}_1 \to\dotsm\to \Vec{x}_n)\text{,}
\end{equation} 
with~$\Vec{x}_0=\Vec{a}$ and~$\Vec{x}_n=\Vec{b}$, and vectors stand for the states of the two-particle state, rather than merely points in space, so this ``path'' corresponds to the two paths in~\Figref{fig:exchangepath}(i).

Here we encounter a problem. As these are identical particles, to each step in that exchange should correspond a term~$\alpha^k_{\textsl{dir}}\pm\alpha^k_{\textsl{op}}$, containing the amplitudes for both~\Figref{fig:exchangepath}~(ii) and~(iii), and the total amplitude should therefore be a product of these terms:
\begin{equation}
\Amp_{\textsl{exch}}=\left(\alpha^1_{\textsl{dir}}\pm\alpha^1_{\textsl{op}}\right)\left(\alpha^2_{\textsl{dir}}\pm\alpha^2_{\textsl{op}}\right)\dotsm\left(\alpha^N_{\textsl{dir}}\pm\alpha^N_{\textsl{op}}\right)\text{;}
\end{equation}
fortunately, as we add more and more intermediate points, the direct transition amplitudes~$\alpha^k_{\textsl{dir}}$ become of the form~$\exp(i\Delta t)$, while an opposite transition will go along a finite path of length~$D$, the distance between the particles over a shorter and shorter time, leading to an amplitude~$\alpha^k_{\textsl{op}}$ of the form~$\exp(L/\Delta t)$, meaning that any term in the sum after multiplying out the terms in~$\Amp_{\textsl{exch}}$ which contains such a contribution would diverge as the time difference between the points goes to zero, in the sense that the phases would cancel each other out. So we end up with:
\begin{equation}
\Amp_{\textsl{exch}}=\alpha^1_{\textsl{dir}}\alpha^2_{\textsl{dir}}\dotsm\alpha^N_{\textsl{dir}}=e^{i\varphi}\exp\left(i\Action\{\Vec{x}_0\to \Vec{x}_1 \to\dotsm\to \Vec{x}_n\}/\hbar\right)
\end{equation}
incorporating the same topological factor as in~\Figref{fig:allexchanges}.

However, there is a final subtlety. The variables~$\Vec{x}_i$ that we use to express the particles' location in space are redundant for identical particles, and we are starting and ending in the same point as far as they are concerned. In order to avoid ambiguity, before making any calculations, we must choose which transitions are direct and which are opposite, which we can do by choosing once and for all a subspace of the two-particle space which is the space of actually measurable parameters. Then the path in~\Figref{fig:exchangepath}(i) passes through the edge of this subspace, and for this transition, (ii) and~(iii) switch roles, so the factor is~$\alpha^{\ell}_{\textsl{op}}\pm\alpha^{\ell}_{\textsl{dir}}$. Such transitions happen only once for a sensible choice of subspace for a direct exchange, so we then get a final phase of~$\pm e^{i\varphi}$, and if we incorporate the operational phase into the angle, that is~$\phi=\varphi$ for operational bosons and~$\phi=\varphi+\pi$ for operational fermions, we get a final phase that is~$e^{i\phi}$, that is, we get anyonic behavior.

This last point ties back into the reduced configuration space approach; but while~\citet*{LeinaasMyrheim1977} started with configuration space reduction and then quantized, opening the door to potential complications in the process of quantization, we took advantage of the ease of making calculations (and using topological reasoning) for distinguishable particles, and then combined them operationally, while, as we have just seen, still allowing for anyons.
\section{What is to be done next?}
So we have found that the operational approach is compatible with anyons, despite initially seeming to exclude them. We also found a way of expressing the anyonic behavior using topology, without particle identity. Both a derivation of the topological improvement underpinning our approach to distinguishable anyons~\cite{NeoriGoyal2014a}, as well as a rigorous treatment of the distinguishable anyons themselves and their reconciliation with the operational approach~\cite{NeoriGoyal2014b}, are in preparation.

\begin{theacknowledgments}
KHN would like to thank Oleg Lunin for long discussions about this research throughout its development, Lawrence S. Schulman for extensive feedback through correspondence, and John Skilling and Marco Varisco for illuminating comments.
\end{theacknowledgments}

\bibliographystyle{aipproc} 
\bibliography{NeoriKlilHResearch}
\end{document}